# The Imaginary Mass of the Neutrino

Jose N. Pecina-Cruz
The University of Texas-Pan American
Department of Physics
1201 West University Drive,
Edinburg, Texas 78541
E-mail: jpecina2@panam.edu

**Abstract**

This paper proposes that the elusive neutrino is a space-like unitary irreducible representation of the Poincare Group. Although its imaginary mass is not a physical observable, its square is. This amount is found experimentally to be negative. Therefore, the neutrino is a tachyon with a measurable property; the square of its mass. This paper does not resolve the problem of how the neutrino propagates in the space. Further studies are necessary to address this question.

**Introduction**

It has been reported by different groups of experimentalists that the mass square of the neutrino is negative [1]. It is the purpose of the present manuscript to suggest a logical explanation to the fact that the square of the mass of the neutrino is negative. The unitary irreducible representations of the Poincare Group were classified by E. Wigner [2], and completed by Bargmann, Niarmak, and Chandra [3] in 1947. Based on the irreducibility of the representations, they identified the unitary irreducible representations as elementary particles. Since the concept of an elementary particle rests on the lack of further structure of the matter at that level. However, Wigner could not find a physical meaning to the representations in the region 1, 3, and 5 of the space-time like cone shown in Figure 1. In this manuscript, the unitary irreducible representations of the Group of Poincare on the region 1 are discussed and identified as the representations that describe the neutrino. In section 1 the space-like unitary irreducible representations of the Poincare Group will be discussed.

**I. The Space-like Unitary Irreducible Representations of the Poincare Group**

In the space-like, region 1, the eigenvalue of the square of the four-momentum is the negative of the square of the rest mass of the particle. Therefore, the mass would be an imaginary number. No physical interpretation of elementary particles had been associated to this region prior to this paper.

| Orbit of the group | Interval | Vector $k_0$ in momentum space | Little Group of $k_0$ |
|---|---|---|---|
| 1. ŝ · ŝ < 0, | space-like | (0, 0, k, 0) | SL(2,R) |
| 2. ŝ · ŝ > 0, t > 0, | future time-like | (0, 0, 0, k) | SO(3) |
| 3. ŝ · ŝ > 0, t < 0, | past time-like | (0, 0, 0, -k) | SO(3) |
| 4. ŝ · ŝ = 0, ŝ ≠ 0, t > 0, | future null | (0, 0, k, k) | ISO(2) |
| 5. ŝ · ŝ = 0, ŝ ≠ 0, t < 0, | past null | (0, 0, k, -k) | ISO(2) |
| 6. ŝ = 0, | the zero vector | (0, 0, 0, 0) | SO(3,1) |

The regions and the little groups of the four-vectors $k_0$ are shown in the table above. Figure 1 shows the space time regions of the representations of the group of Poincare. The time-like and space-like orbits are shown as thick lines; since the Heisenberg uncertainty principle allows tunneling from one to the other, as it is discussed in Ref. 4.

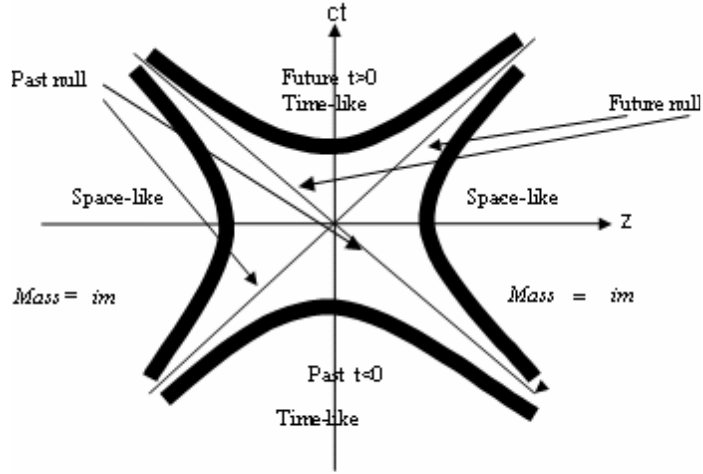

Figure 1

The little group of the four-vector $\hat{k}_0 = (0,0,k,0)$ is the SL(2,R) which contains spin representations. These representations were classified independently by Bargmann, Naimark and Gelfand, and Harish-Chandra[3].
It can be shown that a general basis vector can be constructed as

$$\left|\hat{k}m\right\rangle = T(W_k)\left|\hat{k}_0 m\right\rangle = T(R_z(\theta)T(L_x(p))T(L_z(p))\left|\hat{k}_0 m\right\rangle \qquad (1)$$

Where $L_x(p)$, $L_z(p)$ are Lorentz boost in X and Z direction and $R_z(\Theta)$ is rotation around the axis Z.

The unitary irreducible representations of the Poincare group in the space-like region are obtained by following the technique of the little group. By using Equation (1) these are given by

$$D(W)|\hat{k}m> = \sum_{m'}|\hat{k}m'> D_{m'm}(W_{\hat{k}_0})$$

(2)

$$D(u)|\hat{k}m> = \exp(i\hat{k}.\hat{u})|\hat{k}m>$$

Where $D_{m'm}(W_{\hat{k}_0})$ is a representation of the little group SL(2,R). The vectors, $|\hat{k}m>$, of the representations of Poincare group are labeled by the eigenvalues of its Casimir operators and the general invariant $Q^2$ [5]:

$$p^2 = -k^2 \tag{3}$$

$$Q^2 = \frac{W^2}{p^2} \tag{4}$$

$$Q^2 = (H_{10})^2 + (H_{20})^2 + (H_{12})^2 \tag{5}$$

Where, the general invariant $Q^2$ is the ratio of the Pauli-Lubanski operator $W^2$ and the four-momentum, $p^2$.

Because $Q^2$ commute with all the members of the Lie algebra, it is proportional to the unit matrix, $Q^2=q*1$. The eigenvalues of $H_{12}$ are $m$, which can take integral or half integral values. The space-like representations of the Poincare group are labeled by k, q, and m as it is illustrated by Equation 2. The possible representations $D_{m'm}(W_{\hat{k}_0})$ of SL(2,R) are [3]:

1) $C^0_q$: q > 0, m=0,±1, ±2, ±3, …
2) $C^{1/2}_q$: ¼ < q < α, m=±1/2, ±3/2, …
3) $D^-_k$: k=1/2,1,3/2…; q=k(k-1), m=k, k+1, k+2,…
4) $D^\pm_k$: k=1/2,1,3/2…; q=k(k-1), m=-k, -(k+1), -(k+2),…

$C^0_q$, and $C^{1/2}_q$ are called continues classes because q is continue. $D^\pm_k$ are called discrete classes. In this paper the representations corresponding to the discrete series with m=±1/2 are identified with the neutrino, and m=±1/2 as its helicity. The eigenvalue, $k^2$<0, is identified as the square of the neutrino mass. The mass of the neutrino is imaginary without any physical meaning, but its square. The neutrino is in fact a tachyon

## Conclusion

The fact that the square of the mass of the neutrino and that the space-like unitary irreducible representations of the Poincare group have an imaginary mass strongly suggest that the mass itself does not have an independent physical meaning. This was already pointed out by Ref. 4. The eigenvalue of the four-momentum square is, $p^2 = -m^2$. This suggests that the physical measurable amount is the square of the four-momentum.

## Acknowledgments

I would like to thank Roger Maxim Pecina for going through this article.